\documentclass[12pt]{iopart}

\usepackage{iopams,graphicx}
\begin{document}
\title{Strong-coupling dynamics of Bose-Einstein condensate in a double-well trap}
\author{V.O. Nesterenko$^1$, A.N. Novikov$^1$ and E. Suraud$^2$}
\address{$^1$Bogoliubov Laboratory of Theoretical Physics, Joint
Institute for Nuclear Research, Dubna, Moscow region, 141980,
Russia}
\ead{nester@theor.jinr.ru}
\address{$^2$Laboratoire de Physique Quantique, Universit\'e Paul
Sabatier, 118 Route de Narbonne, 31062 cedex, Toulouse, France}

\begin{abstract}
Dynamics of the repulsive  Bose-Einstein condensate (BEC) in a
double-well trap is explored within the 3D time-dependent
Gross-Pitaevskii equation. The model avoids numerous common
approximations (two-mode treatment, time-space factorization,
fixed values of the chemical potential and barrier penetrability,
etc) and thus provides a realistic description of BEC dynamics,
including both weak-coupling (sub-barrier) and strong-coupling
(above-barrier) regimes and their crossover. The strong coupling
regime is achieved by increasing the number $N$ of BEC atoms and
thus the chemical potential. The evolution with  $N$ of Josephson
oscillations (JO) and Macroscopic Quantum Self-Trapping (MQST) is
examined and the crucial impact of the BEC interaction is
demonstrated. At  weak coupling, the calculations well reproduce
the JO/MQST experimental data. At  strong coupling, with a
significant overlap of the left and right BECs, we observe a
remarkable persistence of the Josephson-like dynamics: the JO and
MQST converge to a high-frequency JO-like mode where both
population imbalance and phase difference oscillate around the
zero averages. The results open new avenues for BEC
interferometry.
\end{abstract}

\submitto{\jpb}
\maketitle

\section{Introduction}
The trapped Bose-Einstein condensate (BEC) is nowadays widely
recognized as a source of  new fascinating physics, see monographs
\cite{Petrick_Smith_02,Pit_Str_03} as well as early
\cite{Dalfovo_99,Leggett_01,Yuk_rew_01} and recent
\cite{Gati_07,Bloch_08} reviews. Among diverse aspects of this
field, a large attention is paid to dynamics of bound condensates
and relevant nonlinear effects caused by the interaction between
BEC atoms \cite{Gati_07,Mele_NJP_11}. A boson Josephson junction
in a double-well trap represents a typical example of  relevant
system. Its basic dynamical regimes, Josephson oscillations (JO)
and Macroscopic Quantum Self-Trapping (MQST), were widely
investigated in the weak coupling limit, both in  theory
\cite{Gati_07,Mele_NJP_11,Smerzi_97,Raghavan_99} and experiment
\cite{Albiez_exp_PRL_05,Gati_APB_06,Levy_Nature_07}. Last years,
new processes in bound condensates, like  controlled BEC transport
\cite{Nest_LP_09,Nest_JPB_09}, matter wave interferometry
\cite{Schumm_Nature_05}, squeezing and entanglement
\cite{Esteve_Nature_08}, were in the focus. Being diverse and
sophisticated, these processes concern, nevertheless, the basic JO
and MQST dynamics and, what is important, often occur beyond the
weak coupling limit. Thus, the study of JO/MQST at strong coupling
(SC), e.g. in the crossover between sub-barrier and above-barrier
transfer, becomes essential.

The SC regime routinely arises at a strong interaction or/and
large number of BEC atoms. Hence  this regime is quite common. At
the same time, its features have not been yet properly
investigated.  Note that at SC, the left and right BECs in a
double-well trap are not well separated, so that  the use of the
ordinary variables of  Josephson dynamics, population imbalance
$z$ and phase difference $\theta$, is not well justified. There
thus arises a general question whether the JO and MQST survive at
SC and, if yes, to what extent they are modified?

Numerous previous studies of  Josephson dynamics in a double-well
trap were  performed within the two-mode approximation (TMA)
\cite{Milburn_97}, where only two lowest energy levels of the
system contribute to the dynamics and a weak coupling of the
condensates is assumed.  Obviously,  SC dynamics can involve many
energy levels and the TMA is then not correct. Even the TMA
modifications, like a variable tunneling model \cite{Ananikian_06}
and a direct implementation of Wannier states
\cite{Ostrovskaya_00}, do not amend the principle TMA limitation
to deal only with two lowest BEC states and thus leave TMA
ineligible to the SC case.

The SC dynamics requires  a more involved theory embracing  impact
of all  excited states and avoiding, as much as possible, other
standard approximations, e.g.  the space-time factorization and
partition (for every well) of the total order parameter. There are
already some studies of this kind
\cite{Gati_07,Mele_NJP_11,Lee_JPB_07,Streltsov_PRL_07,Sakmann_PRL_09}.
Most of them consider transformation of the trap from a double to
a single well shape \cite{Lee_JPB_07} or back
\cite{Streltsov_PRL_07,Sakmann_PRL_09} and thus partly concern the
SC case. The treatment varies from using the Gross-Pitaevskii
equation \cite{GPE} in \cite{Lee_JPB_07} to many-body quantum
dynamics in \cite{Streltsov_PRL_07,Sakmann_PRL_09}. These studies
confirm the important role played by the excited states,  beyond
the TMA. However, these studies do not especially address the
evolution of the Josephson dynamics when approaching the SC and do
not provide a relevant analysis. Moreover, these studies are
usually performed in the one-dimensional (1D) limit
\cite{Olshanii_PRL_98}, the relevance of which is not well
demonstrated for the SC. In addition, the 1D limit is obviously
questionable in a true three-dimensional (3D) case,  which is
often met in practice, see e.g. the JO/MQST Heidelberg experiment
\cite{Albiez_exp_PRL_05,Gati_APB_06}.

In this paper, we analyze the evolution of the JO/MQST dynamics in
a double-well trap while transforming the system from the weak
coupling to the SC. The transfer is achieved by increasing the
number of BEC atoms from $N$=1000 to 10000.  We thus obtain the
rise of the cumulative effect of the repulsive interaction and the
subsequent upshift of the chemical potential.  The analysis is
based on an explicit solution of the 3D time-dependent
Gross-Pitaevskii equation for the total order parameter. None of
the questionable approximations mentioned above is thus  used.
Both the population imbalance $z$  and  phase difference $\theta$
are examined.

The calculations are performed for a double-well configuration and
the JO/MQST initial conditions of the Heidelberg experiment
\cite{Albiez_exp_PRL_05,Gati_APB_06}.  For $N$=1000, we reproduce
the JO/MQST experimental data \cite{Albiez_exp_PRL_05,Gati_APB_06}
for the repulsive condensate of  $^{87}$Rb atoms. This justifies
the relevance of our model in the weak coupling limit. By
increasing the number of atoms up to $N$=10000, we approach the SC
regime and show that, despite an essential intersection of the
left and right BECs in the double-well trap, these BECs still keep
their individuality. Hence, using the relative Josephson conjugate
variables $z$ and $\theta$ remains reasonable.  We show that JO
survive at SC, though with a higher frequency. The MQST is
transformed to a similar JO. Actually, JO and MQST merge to the
same mode. Perhaps, a similar MQST $\to$ JO transfer was earlier
predicted for 1D BEC as a reappearance of tunneling in the strong
interaction limit \cite{Julia_Diaz_10}.

In most of the previous calculations (see, e.g.
\cite{Mele_NJP_11}), the initial conditions are obtained by constraining
the system into a non-stationary state of the symmetric trap. This
 may be questionable at SC.  In this
respect, we build initial conditions within a more
realistic technique  \cite{Albiez_exp_PRL_05,Gati_APB_06}.
Namely, a stationary state with the proper initial conditions is
produced in an asymmetric trap and then the trap is
non-adiabatically transformed to the symmetric form.

The paper is organized as follows.  The calculation scheme is sketched in Sec.
\ref{sec:calc_scheme}. Results of the calculations are discussed in Sec. \ref{sec:results}.
A summary is given in Sec. \ref{sec:summary}.

\section{Calculation scheme}
\label{sec:calc_scheme}

We solve the 3D time-dependent Gross-Pitaevskii equation
\cite{GPE}
\begin{equation}\label{GPE}
 i\hbar\frac{\partial\Psi}{\partial t}({\bf r},t) =
[-\frac{\hbar^2}{2m}\nabla^2 + V({\bf r}) + g_0|\Psi({\bf
r},t)|^2]\Psi({\bf r},t)
\end{equation}
for the total order parameter $\Psi({\bf r},t)$ describing  the BEC in
both left and right wells of the trap. Here $g_0=4\pi\hbar^2 a_s
/m$ is the interaction parameter, $a_s$ is the scattering length,
and $m$ is the atomic mass. The trap potential
\begin{equation} \label{trap_pot}
V({\bf r})=\frac{m}{2}(\omega^2_x x^2+\omega^2_y y^2+\omega^2_z
z^2)+V_0 \cos^2(\pi x/q_0)
\end{equation}
includes the anisotropic harmonic confinement and the barrier in
$x$-direction, where $V_0$ is the barrier height and $q_0$
determines  the barrier width. Note that the barrier parameters
may depend on time at the stage of preparation of the initial
conditions.

Following the experiment \cite{Albiez_exp_PRL_05,Gati_APB_06}, we
use a BEC of $^{87}$Rb atoms with $a_s=5.75$ nm. The trap
frequencies are $\omega_x=2\pi\times 78$ Hz, $\omega_y=2\pi\times
66$ Hz, $\omega_z=2\pi\times 90$ Hz, i.e.
$\omega_y+\omega_z=2\omega_x$
\cite{Albiez_exp_PRL_05,Gati_APB_06}. The barrier parameters are
$V_0=420\times h$ Hz and  $q_0=5.2 \; \mu$m
\cite{Albiez_exp_PRL_05,Gati_APB_06}. The distance between centers
of the left and right wells is then $d=$4.4 \;$\mu$m. For $N$=1000
atoms, we reproduce the conditions of the JO/MQST Heidelberg
experiment \cite{Albiez_exp_PRL_05,Gati_APB_06} for  a weak
coupling of the left and right BEC fractions through the barrier.

The static solutions of (\ref{GPE}) are found within the damped
gradient method \cite{DGM} while the time evolution is computed
within the time-splitting  \cite{TSM} and fast
Fourier-transformation techniques. The order parameter $\Psi({\bf
r},t)$ is determined in the 3D cartesian grid. The requirement
$\int^{-\infty}_{+\infty}dr^3 |\Psi({\bf r},t)|^2 =N$ is directly
fulfilled by using an explicit unitary propagator. Reflecting
boundary conditions are used throughout, but have no impact on the
dynamics because of the harmonic confinement. In both static and
time-dependent cases, only the order parameter with the lowest
energy (chemical potential $\mu$) is explicitly computed. No
time-space factorization is used. The conservation of the total
energy $E$ and number of atoms $N$ is controlled. Note that the
Gross-Pitaevski equation  is mathematically equivalent to the
non-linear Schr\"odinger equation. In this sense, our approach is
a counterpart of the time-dependent Hartree-Fock method for the
system of interacting bosons.

The JO and MQST are studied in terms of the time-dependent
normalized population imbalance $z$ and phase difference $\theta$,
\begin{equation}\label{z_t}
z(t) = \frac{N_L(t)-N_R(t)}{N} \; , \quad \theta (t) =
\phi_R(t)-\phi_L(t) \; ,
\end{equation}
where $N_{L,R}(t)$ are respectively populations of the left and right wells
(with $N_L(t)+N_R(t)=N$) and $\phi_{L,R}(t)$ are the corresponding
BEC phases. The populations read
\begin{equation}\label{N_LR}
N_{j}(t)=\int^{+\infty}_{-\infty}dr^3 |\Psi_{j}({\bf r},t)|^2
\end{equation}
with $j = L, R$ and $\Psi_{L}({\bf r},t)=\Psi(x\le 0,y,z,t)$,
$\Psi_{R}({\bf r},t)=\Psi(x\ge 0,y,z,t)$.

The phases $\phi_{j}(t)$ are defined as
\begin{equation}\label{phi_LR0}
\phi_{j}(t)=\arctan \frac{\gamma_j(t)}{\zeta_j(t)}
\end{equation}
with the averages
\begin{eqnarray}\label{gamma}
\gamma_j (t)=\frac{1}{N_j}\int^{+\infty}_{-\infty}dr^3
\mathrm{Im}(\Psi_j({\bf r},t))|\Psi_j({\bf r},t)|^2 \; ,
\\ \label{zeta}
\zeta_j (t)=\frac{1}{N_j}\int^{+\infty}_{-\infty}dr^3
\mathrm{Re}(\Psi_j({\bf r},t))|\Psi_j({\bf r},t)|^2 .
\end{eqnarray}
Computation of the phase time evolution through $\arctan$ may be
cumbersome. So we use (\ref{phi_LR0}) only for the static case
while the time evolution is calculated through the phase
increments $\phi_{j}(t+\delta t)\approx \phi_{j}(t)+\delta
\phi_{j}(t)$ for a small time step $\delta t$.  Namely, we use
\begin{equation}\label{dphi}
\delta\phi_{j}(t)=
\sqrt{\frac{[\delta\gamma_j(t)]^2+[\delta\zeta_j (t)]^2}
{\gamma_j^2(t+\delta t)+\zeta_j^2(t+\delta t)}}
\end{equation}
with $\delta\gamma_j(t)=\gamma_j(t+\delta t)-\gamma_j(t)$,
$\delta\zeta_j(t)=\zeta_j(t+\delta t)-\zeta_j(t)$.

The calculations are performed for $N$=1000, 3000, 5000, and 10000
atoms. For all the cases, the initial population imbalance  $z_0$
is 0.3 for JO and 0.6 for MQST
\cite{Albiez_exp_PRL_05,Gati_APB_06}. The initial phase difference
is $\theta_0$=0 for both JO and MQST.

In most of the previous studies, the initial state
is prepared as the lowest {\it non-stationary} state
with the {\it constrained}  $z_0$ in
the {\it symmetric} trap, see, e.g. \cite{Mele_NJP_11}. The constraint is
reasonable for a weak coupling ($N$=1000 in our study) but may be
questionable for SC  where the chemical potential $\mu_x$ associated to the
motion in x-direction approaches or even exceeds the barrier height.
Besides, the constraint procedure deviates from the actual experimental
initialization \cite{Albiez_exp_PRL_05,Gati_APB_06} of the JO/MQST dynamics.
Hence we use here, in addition to the constraint calculations, a more reliable
and realistic initialization following \cite{Albiez_exp_PRL_05,Gati_APB_06}.
We start from the asymmetric trap produced from the
symmetric one by a right-shift $d$ of the barrier. The value of
the shift is adjusted to provide the given $z_0$
in the lowest {\it stationary} state of the {\it asymmetric}
trap. In some cases, an additional widening of the barrier, $q_0 \to q$,
in the asymmetric trap is applied. Then the trap is rapidly (for a time $\tau$)
returned to the symmetric form and the JO/MQST time evolution starts.

The parameters of the procedure are given in Table 1. The return
time $\tau$ is chosen to provide a reasonable initialization of
JO/MQST. The calculations show that a too rapid return shakes the
system and leads to a fussy and fragile JO/MQST. Instead, for too
slowe return, the equilibration process noticeably modifies  $z$
and $\theta$ from their initial values $z_0$ and $\theta_0$.
Altogether, this procedure is more justified and  realistic  than
the mere constraint. For $N$=1000, it  fully reproduces formation
of the initial conditions in the experiment
\cite{Albiez_exp_PRL_05,Gati_APB_06}.
\begin{table}
\caption{\label{arttype} Parameters of the procedure
 for the JO/MQST initial conditions. See text for detail.}
\begin{center}
\begin{tabular}{|c|c|c|c|c|c|c|}
\hline
  N &\multicolumn{2}{|c|}{$d$, $\mu$m} &\multicolumn{2}{|c|}{$\tau$, ms}
   &\multicolumn{2}{|c|}{$q$, $\mu$m}\\ \cline{2-7}
    & JO  &  MQST  &JO &MQST & JO & MQST \\
\hline
 1000&0.25&0.5 &4 &8 & 5.2 & 5.2 \\
 \hline
 3000& 0.45 &1.05 &5 &9 & 5.2 & 5.2\\
 \hline
 5000& 0.62 &1.35 &5 &5 & 5.2 & 6.6 \\
 \hline
 10000& 1.35 &1.35 &8 &2 & 5.2  & 10.1 \\
\hline
\end{tabular}
\end{center}
\end{table}

\section{Results and discussion}
\label{sec:results}
\subsection{Static interaction effect}
In our study, we swap from  weak to  strong coupling (SC) and
approach the crossover point by increasing the number of atoms $N$
from 1000 to 10000. This results in rising the integral interaction effect
and a subsequent growth of the chemical potential $\mu_x$. Thus
we naturally come from  deeply sub-barrier to above barrier cases.
\begin{figure}[h]
\begin{center}
\includegraphics[width=10cm]{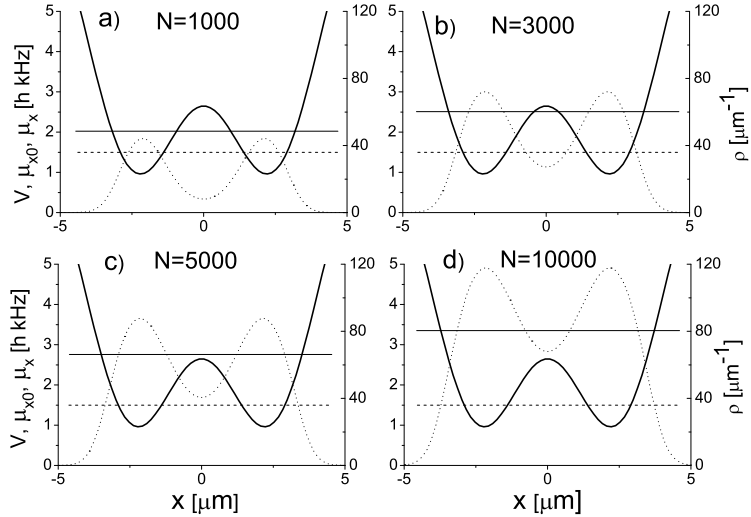}
\end{center}
\caption{The double-well trap potential $V(x)$ (bold curve), the
linear (without interaction) chemical potential $\mu_{0x}$ (dotted
line), the non-linear (with interaction) chemical potential
$\mu_{x}$(solid line), and BEC density $\rho(x)$ (dotted curve) in
the stationary states with $N_1=N_2=N/2$ for $N$=1000 (a), 3000
(b), 5000 (c), and 10000 (d). }
\end{figure}

This is demonstrated in Fig. 1 where the static BEC density $\rho (x)$ and chemical
potential $\mu_x$ (for the motion in x-direction) are exhibited at
different values of $N$ and compared to the relevant trap potential
$V_x(x)=m\omega^2_x x^2/2+V_0 \cos^2(\pi x/q_0)$. As mentioned
above, the trap parameters are the same as in the experiment
\cite{Albiez_exp_PRL_05,Gati_APB_06}. The density
\begin{equation}
\rho (x)=\int^{+\infty}_{-\infty} dy dz |\Psi (x,y,z)|^2
\end{equation}
is determined from the solutions of (\ref{GPE}) for interacting
BEC.

Figure 1 shows that, for $N$=1000, the overlap of the left and
right BECs is small. The ratio of the densities  at their maxima
($x=\pm 2.2 \;\mu$m) and center of the trap $(x=0)$ is
$\rho_m/\rho_c$=5.1. This is the case of the weak coupling used in
the experiment \cite{Albiez_exp_PRL_05,Gati_APB_06}. Increasing
$N$ results in rising the densities and larger overlap in the
barrier region. At $N$=10000, we already have the SC case with
$\rho_m/\rho_c$=2. Here the overlap of the left and right BECs
cannot be neglected and the system should be treated with the {\it
total} order parameter. At the same time, the left and right
density bumps are still well distinctive. So a physical view of
the system in terms of two (strongly coupled) BECs is still
reasonable and one may yet expect for the JO/MQST dynamics
described by the relative variables $z$ and $\theta$.

To discriminate the sub-barrier and above-barrier cases, one
should compare the chemical potential
 $\mu_x$ and the barrier height $V_0$. For $\mu_x$, only the motion in $x$-direction has to be
 taken into account. In the linear case ($g_0$=0),  we straightforwardly get
\begin{equation}\label{mu_x0}
\mu_{x0} = \mu_0 - \frac{\hbar}{2}(\omega_y+\omega_z)
\end{equation}
Since $\omega_y+\omega_z=2\omega_x$
\cite{Albiez_exp_PRL_05,Gati_APB_06}, the subtractive term in
(\ref{mu_x0}) is equal to $\hbar \omega_x$. The calculations give
$\alpha =\mu_{x0}/ \mu_0=3/4$, i.e. just $x$-motion mainly
contributes to the total chemical potential  $\mu_0$. This is
because the barrier separates the harmonic $x$-confinement into
two more narrow regions and thus effectively increases $\omega_x$.

In the nonlinear case ($g_0 \ne$ 0), the estimation of $\mu_x$  is
straightforward for 1D system but complicated for 3D one. Here we
roughly put $\mu_x = \alpha \mu$ where $\mu$ is the total {\it
nonlinear} chemical potential. Hence we suppose that in the linear
and nonlinear cases the relative contribution of x-motion to the
chemical potential is the same.

Figure 1 compares the linear $\mu_{x0}$ and nonlinear $\mu_x$
to the barrier height $V_0$. It is seen that $\mu_{x0}$ does not
depend on $N$. It is always much lower than $V_0$,
thus leading  to the deeply sub-barrier case. If the repulsive
interaction is switched on, the nonlinear chemical potential $\mu_x$ rises
with $N$. We see the sub-barrier case for $N=1000$, the crossover region for $N$=3000-5000,
and the above-barrier case for $N$=10000.

\subsection{JO and MQST evolution}

The evolution of JO and MQST with $N$ is demonstrated in Figs.
2-3. In all  cases, the initial ($t$=0) conditions  are $z_0=$0.3,
$\theta_0$=0 for JO and $z_0=$0.6, $\theta_0$=0 for MQST
\cite{Albiez_exp_PRL_05,Gati_APB_06}. Both the constraint
technique (CT) and barrier-shift technique (BST)
\cite{Albiez_exp_PRL_05,Gati_APB_06}, described in Sec.
\ref{sec:calc_scheme}, are used for initialization of the
dynamics. For the BST, the time $\tau$ when the asymmetric trap is
fully reduced to the symmetric form  is marked by a vertical line.
It is seen that CT at $t$=0 and BST at $t=\tau$ give somewhat
different $z$ and $\theta$. The larger $N$ and $\tau$ (and thus
the equilibration time), the more the difference. Nevertheless,
the CT and BST usually initiate a similar (up to a constant time
shift) dynamics, especially for JO. In what follows, we will
mainly exam the BST results.
\begin{figure}[h]
\begin{center}
\includegraphics[width=13cm]{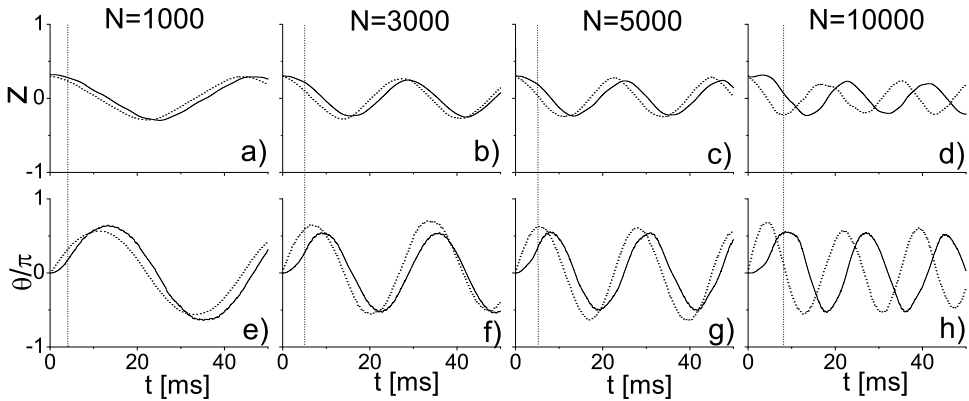}
\caption{ Time evolution of JO population imbalance $z$ (upper
plots) and phase difference $\theta$ (bottom plots) for N=1000,
3000, 5000, and 10000, as indicated. The barrier-shift (solid
line) and constraint (dotted line) techniques are used to initiate
the evolution. For the barrier-shift case, the return time $\tau$
is marked by the vertical dash line. See text for more detail.}
\end{center}
\end{figure}

First of all, note that for $N$=1000 our calculations well reproduce the JO and
MQST experimental data \cite{Albiez_exp_PRL_05,Gati_APB_06}.  Following Fig. 2 a),e),
we obtain for JO the robust  z- and $\theta$-oscillations with the frequency
$\omega_{\mathrm{JO}}=2\pi \times $23 Hz which is close to the
experimental value $\omega^{\mathrm{exp}}_{\mathrm{JO}}=2\pi\times$25 Hz. In Fig. 3a) for MQST,
the calculations give  $z$-oscillations around $\langle z \rangle$=0.6 with the frequency
$\omega_{\mathrm{MQST}}=2\pi \times $72 Hz close to the
experimental value $\omega^{\mathrm{exp}}_{\mathrm{MQST}}=2\pi \times $78 Hz.
 A small underestimation of   the experimental frequencies can be caused by using
 in our calculations a smaller number of atoms, $N$=1000 instead of
 $N \approx 1150 \pm 150$ in the experiment.
In Fig. 3e) for  MQST,  the computed $\theta$  linearly rises with the rate
 $\dot{\theta}= 2\pi \times$ 75 Hz$\approx \omega_{\mathrm{MQST}}$ similar to
the experimental rate $2\pi \times $78 Hz. The calculated
oscillation amplitudes $\Delta
z=z_{\mathrm{max}}-z_{\mathrm{min}}$=0.6, $\Delta
\theta=\theta_{\mathrm{max}}-\theta_{\mathrm{min}}$=1.4$\pi$ for
JO and  $\Delta z$=0.3 for MQST also reproduce the experimental
data. A good agreement of our results with the experiment
\cite{Albiez_exp_PRL_05,Gati_APB_06} proves high accuracy and
realistic character of our method and justifies its application to
the more sophisticated cases considered below.

The evolution of JO with $N$ is exhibited in Fig. 2. It is seen
that, despite a significant change in the conditions from $N$=1000
(weak coupling, sub-barrier transfer) through $N$=3000-5000
(significant coupling, crossover region) to $N$=10000 (strong
coupling, above-barrier transfer), the JO keep the main features:
oscillations of $z$ and $\theta$ with the same frequency
$\omega_{\mathrm{JO}}$ around zero average values. The time shift
between $z$ and $\theta$ is about one-half a period. Amplitudes of
the oscillations do not change with $N$. The evolution with $N$
(or similarly with the interaction $U \sim g_0$ between BEC atoms)
is mainly reduced to a growth of the frequency
$\omega_{\mathrm{JO}}$. This trend is  natural since the larger
$N$, the higher the chemical potential $\mu_{x}$ and the larger
the barrier penetrability $K$. Furthermore, the larger $K$, the
higher $\omega_{\mathrm{JO}}$, as it should be in Rabi-like
oscillations.  Altogether we see that JO survive (with a higher
frequency) even at SC and above-barrier transfer. The crossover
region $N$=3000-5000 is passed monotonically.

Note that the above JO evolution is also supported by the weak-coupling arguments \cite{Smerzi_97}
though application of these arguments needs a word of caution. In the TMA weak-coupling picture for the interacting BEC
\cite{Smerzi_97},
 JO dynamics is driven by the interaction/coupling ratio $\Lambda = NU/K$. With increasing $N$,
the interaction part $NU$ grows linearly while the barrier penetrability $K$
rises exponentially. Altogether, $\Lambda$ falls with $N$ which, following  TMA
calculations \cite{Smerzi_97}, should lead to decreasing $\omega_{\mathrm{JO}}$.  Instead,
our calculations demonstrate the opposite trend. The point is that our approach
is a counterpart of the time-dependent Hartree-Fock method for boson systems, where
the many-body problem for interacting bosons is reduced to one-body problem for a motion
of a boson in an effective one-body potential involving the impact of the interaction.
So the weak coupling arguments \cite{Smerzi_97} should be used in the noninteracting
limit where $\omega_{\mathrm{JO}} \propto K$. Hence the JO trend in Fig. 2.

In Figure 3, the evolution of MQST with $N$ is demonstrated. As
compared to JO, this evolution is more complicated and needs more
time to be exhibited. Hence we use here the larger time interval
80 ms. Fig. 3 shows that MQST is transformed with $N$ to JO from
Fig. 2. Namely, at  $N$=1000, there is an ordinary MQST in
agreement with the experiment
\cite{Albiez_exp_PRL_05,Gati_APB_06}. At $N$=3000-5000,
$z$-oscillations around $\langle z \rangle$=0.6 are gradually
reduced to slower oscillations around $\langle z \rangle$=0 and,
at $N$=10000, basically converge  to JO in Fig. 2d). The linear
evolution of $\theta$ is turned into JO-like oscillations around
the average $\langle \theta \rangle$=2$\pi$. Since the  shift
2$\pi$ is irrelevant,  one actually gets the JO in Fig. 2h).
\begin{figure}[h]
\begin{center}
\includegraphics[width=13.5cm]{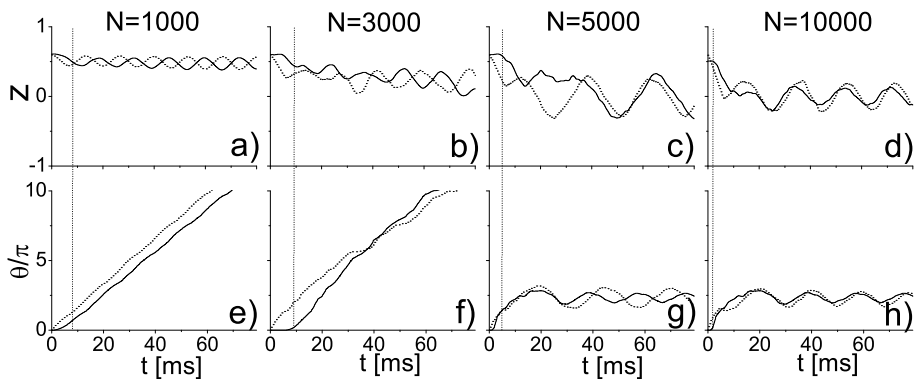}
\caption{The same as in Fig. 2 but for MQST dynamics.}
\end{center}
\end{figure}

A similar MQST evolution was earlier obtained in the strong
interaction limit for 1D BEC \cite{Julia_Diaz_10}. It was shown
that at sufficiently strong interaction the amplitude of MQST
$z$-oscillations starts to grow with $NU_{\mathrm{1D}}$, thus
leading to reappearance of tunneling between two wells. This
effect was explained, within the modified TMA, by coupling the
second and third modes. Our model takes into account all the modes
and thus should cover the result \cite{Julia_Diaz_10} but now for
3D system. Actually, the   transfer MQST $\to$ JO in our
calculations exhibits the similar effect: increasing the amplitude
of $z$-oscillations and thus reappearance of the tunneling.

The most remarkable result of the present study is that, despite a
significant overlap and strong  coupling in the SC case, the left
and right BECs in the trap still keep their individuality and
accept the Josephson-like dynamics in terms of  the relative
variables, population imbalance $z$  and phase difference
$\theta$. In particular, the JO dynamics obtained at a modest
initial $z_0$ remains very regular and robust. This means that BEC
interferometry and related processes may be successfully realized
not only at  weak coupling with a deeply sub-barrier transfer but
also at SC with above-barrier transfer.

\section{Summary}
\label{sec:summary}

The dynamical evolution of  coupled Bose-Einstein condensates (BEC) in a double well trap
was investigated while modifying
 the system from a weak coupling case (small overlap of BECs, deeply
sub-barrier transfer) to a strong coupling case (considerable
overlap of BECs, above-barrier transfer). The evolution was driven
by increasing the number $N$ of BEC atoms and thus rising the
total effect of the interaction between BEC atoms. The numerical
analysis was performed by solving the 3-dimensional time-dependent
Gross-Pitaevskii equation. Thus the two-mode and many other
ordinary approximations were avoided. The main dynamical regimes,
the Josephson oscillations (JO) and Macroscopic Quantum
Self-Trapping (MQST), were inspected.

   The calculations show that  the JO successfully survive even at  strong coupling but acquire a higher frequency.
The MQST is destroyed and finally reduced to the same, though more fragile, high-frequency JO mode.
Altogether we see  that the Josephson-like dynamics certainly persists and remains robust at  strong coupling.
This means that, despite a strong overlap, the left and right BECs in the trap still keep their individuality
and the relative variables, population imbalance $z$  and phase difference $\theta$, remain reliable.
These findings show that  BEC interferometry and related phenomena may be extended
to the case of  strong tunneling coupling, which opens a new avenue for further explorations.

\ack The work was partly supported by the grants  11-02-00086à
(RFBR, Russia), Universit{\'e} Paul Sabatier (Toulouse, France,
2009), and CNRS (2011).  We are grateful to M. Mel\'e-Messenger
for useful discussions.

\end{document}